\documentclass[twocolumn,preprintnumbers,amsmath,amssymb,pra]{revtex4}

\usepackage[colorlinks,linkcolor=blue,anchorcolor=blue,citecolor=blue]{hyperref}
\usepackage{txfonts}
\usepackage{graphicx,hyperref}
\usepackage{subfigure}
\usepackage{dcolumn}
\usepackage{bm}
\usepackage{braket}
\usepackage{upgreek}
\usepackage{extarrows}
\usepackage{setspace}
\usepackage{float}

\begin{document}
	\title{Ratchet current in a $\mathcal{PT}$-symmetric Floquet quantum system with symmetric sinusoidal driving}
\author{Zhiqiang Li$^{1}$}
\author{Xiaoxiao Hu$^{1}$}
\author{Jinpeng Xiao$^{2}$}
\author{Yajiang Chen$^{1}$}
\author{Xiaobing Luo$^{1,2}$}
\altaffiliation{Corresponding author: xiaobingluo2013@aliyun.com}
\affiliation{$^{1}$Department of Physics, Zhejiang Sci-Tech University, Hangzhou, 310018, China}
\affiliation{$^{2}$School of Mathematics and Physics, Jinggangshan University, Ji'an 343009, China}
\date{\today}
\begin{abstract}
	We consider the ratchet dynamics in a $\mathcal{PT}$-symmetric Floquet quantum system with symmetric temporal (harmonic) driving. In the exact $\mathcal{PT}$ phase, for a finite number of resonant frequencies, we show that the long-lasting resonant currents can be generated with the symmetric time-continuous driving, which would otherwise forbid the generation of directed currents in the Hermitian limit. Such a non-Hermitian resonant current can be enhanced by increasing the non-Hermitian level, and in particular, the resonant current peaks (reaches the largest negative
	value) under the condition that the imaginary part of the potential depth is equal to the real part, at which the stable asymptotic current occurs owing to exceptional points (EPs) mechanism. Moreover, the directed currents originating from the symmetry breaking are reported, which increase linearly with the driving frequency, the mechanism behind which is that the cutoff of the momentum eigenstates for the Floquet state with maximum imaginary quasienergy increases as the driving frequency is continuously increased. We also present a non-Hermitian three-level model that can account for the resonant currents and gives surprisingly good agreement with direct numerical results for weak driving, even in the $\mathcal{PT}$-broken regime for the first-order resonance.  Our results provide a new means of realizing the non-Hermiticity-controlled ratchet current by means of a smooth continuous driving, previously used only to generate currents in Hermitian systems.
\end{abstract}
\maketitle

\section{Introduction}
The ratchet effect, that is, directed transport under a zero mean force,  has attracted a continuous interest over the past few decades. This is due to its fundamental and practical importance: on the one hand, its mechanism is relevant to the understanding of quantum chaos and quantum-classical correspondence \cite{J. Gong,DenisovS,S. Denisov,Tobias Salger,Sergey Denisov}, and on the other hand, it has found diverse
applications in many fields, from mechanical devices to quantum systems \cite{Y. V. Gulyaev,P. Hanggi,A. Hashemi,R. W. Simmonds,Y.-J. Wang,B. T. Seaman}. To achieve the ratchet effects, the system must be driven out of equilibrium, and relevant spatiotemporal symmetries, which would otherwise prevent the formation of a directed current, must be broken \cite{S. Flach,P. Reimann,DenisovS,Sergey Denisov}. Besides the classical ratchets, the ratchet phenomenon has also been extended to the quantum regime as it is of great importance, for example, in the design of coherent nanoscale devices \cite{Peter Hanggi,Aref Hashemi}. Due to the high degree of quantum control, Bose-Einstein condensates (BECs) of dilute gases loaded into optical lattices have proven to be excellent candidates for the study of such coherent ratchet effects. Considerable progress has been made in this direction through the study of the quantum kicked rotor \cite{Emil Lundh,Mark Sadgrove,Anatole Kenfack,D. H. White,Jiating Ni,Clement Hainaut}, a paradigm of quantum chaos, which can be realized by exposing a sample of cold atoms to short pulses of an optical standing wave \cite{F. L. Moore}. Other schemes and proposals to implement directed (ratchet) quantum transport in BECs involve the use of time-continuous driving \cite{DenisovS,Tobias Salger,Sergey Denisov,M. Heimsoth}, which has the advantage of being less heating than kick-type driving \cite{M. Heimsoth}. Directed transport in the presence of nonlinearity arising from the many-body nature of BECs has also been investigated in the literature \cite{Dario Poletti,Wen-Lei Zhao,T. S. Monteiro,L Morales-Molina and S Flach,Dario Poletti2,C. E. Creffield,M. Heimsoth}, where it has been found that the interactions between atoms can be harnessed  to generate directed transport even when time-reversal symmetry holds \cite{Dario Poletti}.

So far, most investigations of the quantum ratchet effect have been performed in the context of Hermitian dynamics. The Hermiticity requirement of a Hamiltonian guarantees  its real energy eigenvalues and the conserved total probability. However, a large class of non-Hermitian Hamiltonians that are parity-time symmetric ($\mathcal{PT}$-symmetric) can still have completely real eigenvalue spectra as long as they commute with the parity-time operator \cite{Carl M. Bender,J. K. Boyd,Carl M. Bender2,Carl M. Bende3,S. Weigert,Konstantinos G. Makris,A A Zyablovsky}. One of the most interesting results of such $\mathcal{PT}$-symmetric non-Hermitian Hamiltonian systems is the phase (symmetry breaking) transition, where the spectrum changes from all real (the exact $\mathcal{PT}$ phase) to complex (the broken $\mathcal{PT}$ phase) when the non-Hermitian parameter exceeds a certain threshold \cite{Carl M. Bender,J. K. Boyd,Carl M. Bender2,Carl M. Bende3}. At the transition, the system has exceptional points, also called non-Hermitian degeneracies, at which both the eigenvalues and the eigenvectors of the underlying system coalesce, giving rise to many counterintuitive phenomena \cite{Doppler,Qiongtao Xie,Dashiell Halpern,Parto,Mohammad-Ali,Hodaei,Chong Chen,Bo Zhen,Zin Lin,Feng,W D Heiss}. The study of $\mathcal{PT}$ -symmetric systems has recently been extended to systems with periodically driven potentials (Floquet systems), where the quasienergy (Floquet) spectrum takes the place of the energy spectrum of the static systems \cite{C. T. West,N. Moiseyev,X. Luo,Y. N. Joglekar,Y. Wu,X. B. Luo,M. Chitsazi,Z. Turker,J. Li,B. Zhu,L. Zhou}. A prototype example is the $\mathcal{PT}$-symmetric kicked rotor (KR), where chaos has been shown to assist the exact $\mathcal{PT}$ phase \cite{C. T. West}. The transport properties in such $\mathcal{PT}$-symmetric KR have also attracted some attention in recent years \cite{Stefano Longhi,Wen-Lei Zhao2,J.-Z. Li}. For example, a type of non-Hermitian unidirectional transport, called non-Hermitian accelerator modes, has been discovered in the delocalized regime (quantum resonances) for a $\mathcal{PT}$ extension of the KR model \cite{Stefano Longhi}. Additionally, directed momentum current induced by the $\mathcal{PT}$-symmetric driving in the KR model has been investigated \cite{Wen-Lei Zhao2}.
More recently, Ref. \cite{G. Lyu} has proposed a scheme for generating a persistent current using a static non-Hermitian ratchet. However, knowledge of the quantum ratchet effect in $\mathcal{PT}$-symmetric Floquet systems with time-continuous driving (i.e. a potential that varies smoothly with time rather than being pulsed) remains unknown and needs to be investigated.

In this paper, we investigate the ratchet currents in a $\mathcal{PT}$-symmetric Floquet system with symmetric time-continuous driving. In the exact $\mathcal{PT}$ phase, although the system is conservative on average, we find that the persistent ratchet current can be generated for a finite set of resonant driving frequencies. Such a resonant ratchet current is a hallmark of non-Hermitian transport, as it would disappear in the Hermitian limit. This means that to generate such a non-Hermitian resonant current,  no sawtooth-like asymmetric temporal driving is required. We show that the resonant current  can be enhanced by increasing the non-Hermitian level, and in particular, the time averaged current peaks (reaches a maximum) when the imaginary part of the potential depth is equal to the real part, at which the stable asymptotic current exists due to exceptional points (EPs) mechanism. It is found that the maximum value of the non-Hermitian resonant current  exists  at the EPs for a series of discrete frequencies, i.e. the value of the driving frequencies equal to a half-integer or an integer, which can either correspond to the $\mathcal{PT}$ symmetry breaking points or, surprisingly, arise as an isolated parameter point deep in the $\mathcal{PT}$-unbroken regime. A non-Hermitian three-level model can be used to capture the resonant current, and gives a surprisingly good agreement with the numerically exact results for the weak driving. In the $\mathcal{PT}$-broken regime, it is shown that the asymptotic current increases linearly with the driving frequency. The underlying physics is that, after sufficient evolution time, the system is dominated by the Floquet states whose quasienergy has the largest imaginary part, and the cutoff of the momentum mode of this Floquet state also increases with increasing driving frequency. Finally, the nonlinearity effects on the directed current at EPs are also discussed.

The paper is organized as follows. In Sec. \ref{II}, we describe the model system and show that the introduction of non-Hermiticity gives rise to the Floquet states with asymmetric momentum distribution. In Sec. \ref{III} we study the non-Hermitian resonant currents both numerically and analytically. In Sec. \ref{IV}, the directed currents due to EPs mechanism are identified. In Sec. \ref{V},  the directed currents originating from the symmetry breaking are reported, which are found to increase linearly with the driving frequency. In Sec. \ref{VI},  we show that the non-Hermitian directed current can be induced or suppressed by introducing nonlinearity. The results are summarized in Sec. \ref{VII}.

\section{ Model and current-carrying Floquet eigenstates}\label{II}

We consider condensed atoms confined in a toroidal trap, where the  thickness $r$ of the toroidal trap  is much smaller than its radius $R$, so that lateral motion is neglected and the system is essentially one-dimensional. Hence our problem is described by the dimensionless nonlinear Gross-Pitaveskii (GP) equation (taking $\hbar=m=1$),
\begin{equation}\label{eq1}
	i\frac{\partial}{\partial t}\psi(x,t)=\bigg[\frac{\hat{p}^2}{2}+g|\psi(x,t)|^2+V(x,t)\bigg]\psi(x,t),
\end{equation}
where $x$ is the coordinate, $\hat{p} = -i\partial/\partial x$ is the angular-momentum operator, $g=8NaR/r$ is the scaled strength of nonlinear interaction, $N$ is the number of atoms, and $a$ is  the s-wave scattering length for elastic atom-atom collisions. The condensate is driven by a complex external potential which is periodic in time and has a zero mean, reading
\begin{equation}\label{eq2}
	V(x,t)=K(\sin{x}+i\uplambda\cos{x})\sin{(\omega t+\phi)},
\end{equation}
where $K$ and $\omega$ denote the strength and angular frequency of the flashing potential, $\phi$ is the initial phase of the driving (in our investigation, we will set $\phi=0$. Introducing the initial phase can be used to change the plus-minus sign of the mean currents, thereby steering the transport direction),  and $\uplambda>0$ is the non-Hermitian parameter measuring the strength of the imaginary part of the potential. Throughout this paper, we measure all energies in units of $\hbar^2/2mR^2$, and our analysis will focus mostly on the linear case ($g=0$) unless explicitly stated otherwise.
One can easily verify that the considered Hamiltonian is $\mathcal{PT}$ symmetric, since the system is invariant under the combined action of the parity operator $\hat{\mathcal{P}}: x\to-x$ and the time inversion operator $\hat{\mathcal{T}}: i\to-i$, $t\to-t+2t_0$, where $t_0$ is an appropriate time point.

In Eq. (\ref{eq1}), the  periodic boundary condition, $\psi(x,t)=\psi(x+2\pi,t )$, allows us to expand the wave function as $\ket{\psi(t)}=\sum_{n=-\infty}^{\infty}c_n(t)\ket{n}$, where $\ket{n}=\frac{1}{\sqrt{2\pi}}e^{inx}$ denotes an eigenstate of the undriven
Hamiltonian with (quantized) momentum $n\hbar$.

The time evolution of a quantum state over one period $T = 2\pi/\omega$ is governed by the Floquet operator,
\begin{equation}\label{eq3}
	\hat{U}(t+T,t)=\mathsf{\hat{T}}e^{-i\int_{t}^{t+T}dt\hat{H}(t)},
\end{equation}
where $\hat{\mathsf{T}}$ stands for the time-ordering operator. The eigenequation of the Floquet operator reads
\begin{equation} \label{eq4}
	\hat{U}(t+T,t)\ket{\phi_\alpha(t)}=e^{-i\varepsilon_\alpha T}\ket{\phi_\alpha(t)},
\end{equation}
where $\varepsilon_\alpha$ indicates the quasienergy and $\ket{\phi_\alpha(t)}$ the Floquet mode. As in the undriven case, the Floquet $\mathcal{PT}$ symmetric system is said to be in the unbroken $\mathcal{PT}$ phase if the quasienergies are entirely real, while it is said to be in the broken $\mathcal{PT}$ phase if complex conjugate quasienergies emerge when $\uplambda$ exceeds a certain threshold, i.e., $\uplambda>\uplambda_c$.  Such a  phase transition is referred to as $\mathcal{PT}$-symmetry breaking.

\begin{figure}[htp]	
	\centering
	\includegraphics[width=7cm]{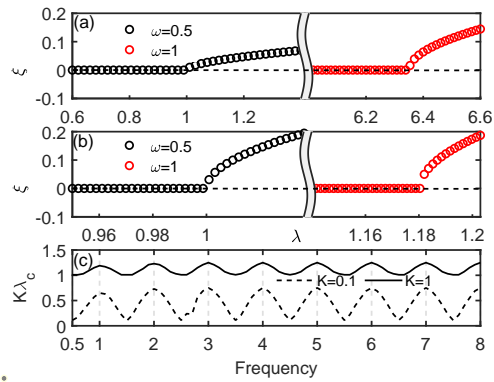}
	\caption{The sum of the absolute values of the imaginary parts of the quasienergies, $\xi=\sum_{\alpha=1}^{N}|\varepsilon_\alpha^i|$, versus the non-Hermitian parameter $\uplambda$ for (a): weak driving $K=0.1$; (b): strong driving $K=1$. Here $\xi$ is shown for two driving frequencies $\omega = 0.5$ (red circles) and $\omega = 1$ (black circles) for each $K$. For $\omega=0.5$, $\mathcal{PT}$ -symmetry breaking is clearly observed at $\uplambda_c=1$ for both weak and strong driving. However, for $\omega = 1$, the $\mathcal{PT}$ symmetry breaking points are different for weak ($\uplambda_c=6.34$) and strong ($\uplambda_c=1.181$) driving. (c): Threshold $K\uplambda_c$ versus driving frequency $\omega$ for a weak driving ($K = 0.1$) and a strong driving ($K = 1$). In the numerical calculation, $\xi$ is summed up over the $N = 512$ Floquet modes.} \label{fig1}
\end{figure}

To identify such a spontaneous $\mathcal{PT}$ symmetry breaking, we numerically compute the sum of the absolute values of the imaginary parts of all quasienergies (see Fig. \ref{fig1}),
\begin{equation}\label{eq5}
	\xi=\sum_{\alpha=1}^{N}|\varepsilon_\alpha^i|,
\end{equation}
where the Floquet matrix is truncated with $N=512$ in momentum space, and $\varepsilon_{\alpha}^i$ denotes
the imaginary part of the $\alpha$-th  quasienergy. Fig. \ref{fig1} shows the numerical behavior of the sum of $|\varepsilon^i_\alpha|$ as a function of the non-Hermitian parameter $\uplambda$ for weak driving $K = 0.1$ [Fig. \ref{fig1} (a)] and strong driving $K = 1$ [Fig. \ref{fig1} (b)]. There is indeed a threshold $\uplambda_c$ above which the sum of $|\varepsilon^i_\alpha|$ increases from zero to a non-zero value, representing a spontaneous $\mathcal{PT}$ symmetry breaking, as shown in Figs. \ref{fig1} (a) and (b).  Additionally, the $\mathcal{PT}$ symmetry breaking threshold varies with the driving frequency. The dependence of the $\mathcal{PT}$ symmetry breaking threshold on the driving frequency is also plotted in Fig. \ref{fig1} (c), where we observe that the $\mathcal{PT}$ symmetry breaking threshold varies almost periodically with the driving frequency, i.e., it reaches a maximum when $\omega$ takes integer values and instead reaches a minimum when $\omega$ takes half-integer values. We also observe that the threshold $K\uplambda$ for $\mathcal{PT}$ symmetry breaking increases with the driving strength $K$.

\begin{figure}[htp]	
 	\center
 	\includegraphics[width=7cm]{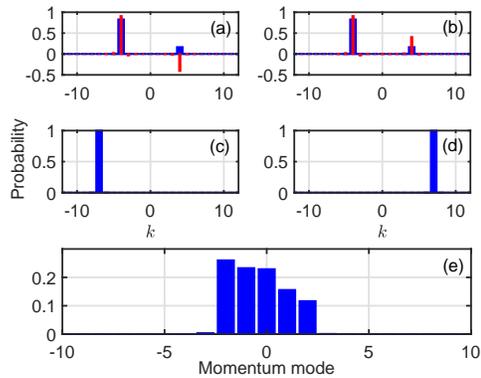}
 	\caption{Occupation of  the momentum modes for three types of Floquet states in the $\mathcal{PT}$-unbroken regime. The blue bars are for the population probabilities in different momentum modes, and the thin red bars [see panels (a) and (b)] are for the distribution of the real parts of the amplitudes in different momentum modes for corresponding Floquet states. (a) and (b) correspond to a pair of degenerate Floquet states with the same momentum occupation, but with different real parts of the probability amplitudes. (c) and (d) are a pair of degenerate Floquet states that occupy the momentum modes in a symmetric manner. These two types have no contribution to the current because of no projection onto the initial state $\ket{0}$. (e): A typical example of a non-degenerate Floquet state that carries a current and overlaps with the initial state. The parameters are $K = 1,\omega = 1$, and $\uplambda = 0.1$.
 	} \label{Fig2}
 \end{figure}

Let us have a look at the properties of the Floquet state that shed light on the understanding of the transport phenomenon. To better distinguish between non-Hermitian and Hermitian situations, we first concentrate on the Floquet states in the unbroken $\mathcal{PT}$ phase, where the quasienergy spectrum is entirely real. In this case, our numerical results show that the Floquet states fall into two classes, degeneracy and non-degeneracy. The momentum distributions of these two typical Floquet states are shown in Fig. \ref{Fig2}. These degenerate Floquet states appear as doublets, which are of two kinds: the pair of degenerate Floquet states that occupy the same momentum component but with different real parts (marked by thin red bars), as shown in Fig. \ref{Fig2} (a)-(b), and the pair of degenerate Floquet states that occupy the negative and positive momentum components in a symmetric manner, as shown in Fig. \ref{Fig2} (c)-(d). In this paper, we assume that the system is initially prepared in the zero-current state $\ket{0}$, which is experimentally convenient because it is the ground state of the undriven Hamiltonian. With such an initial preparation, none of these degenerate states will contribute to the current, since they are well localized in the momentum space and have no projection onto the initial state. Furthermore, our numerical results reveal that all non-degenerate Floquet states have asymmetric momentum distributions around the mode $\ket{0}$, thus acquiring non-zero mean momentum and becoming transporting. A representative example of such a non-degenerate Floquet state is illustrated in Fig. \ref{Fig2} (e). Due to their asymmetric nature, it is reasonable to expect that these non-degenerate Floquet states would give origin to the directed currents of the Floquet $\mathcal{PT}$ symmetry system in the unbroken $\mathcal{PT}$ phase. Finally, we should point out that in the Hermitian limit $\uplambda\to0$, where directed current generation is forbidden, these asymmetric Floquet states can no longer exist.

\section{Non-Hermitian resonant currents}\label{III}
\begin{figure*}[htbp]%
\subfigure{\centering\includegraphics[scale=0.35]{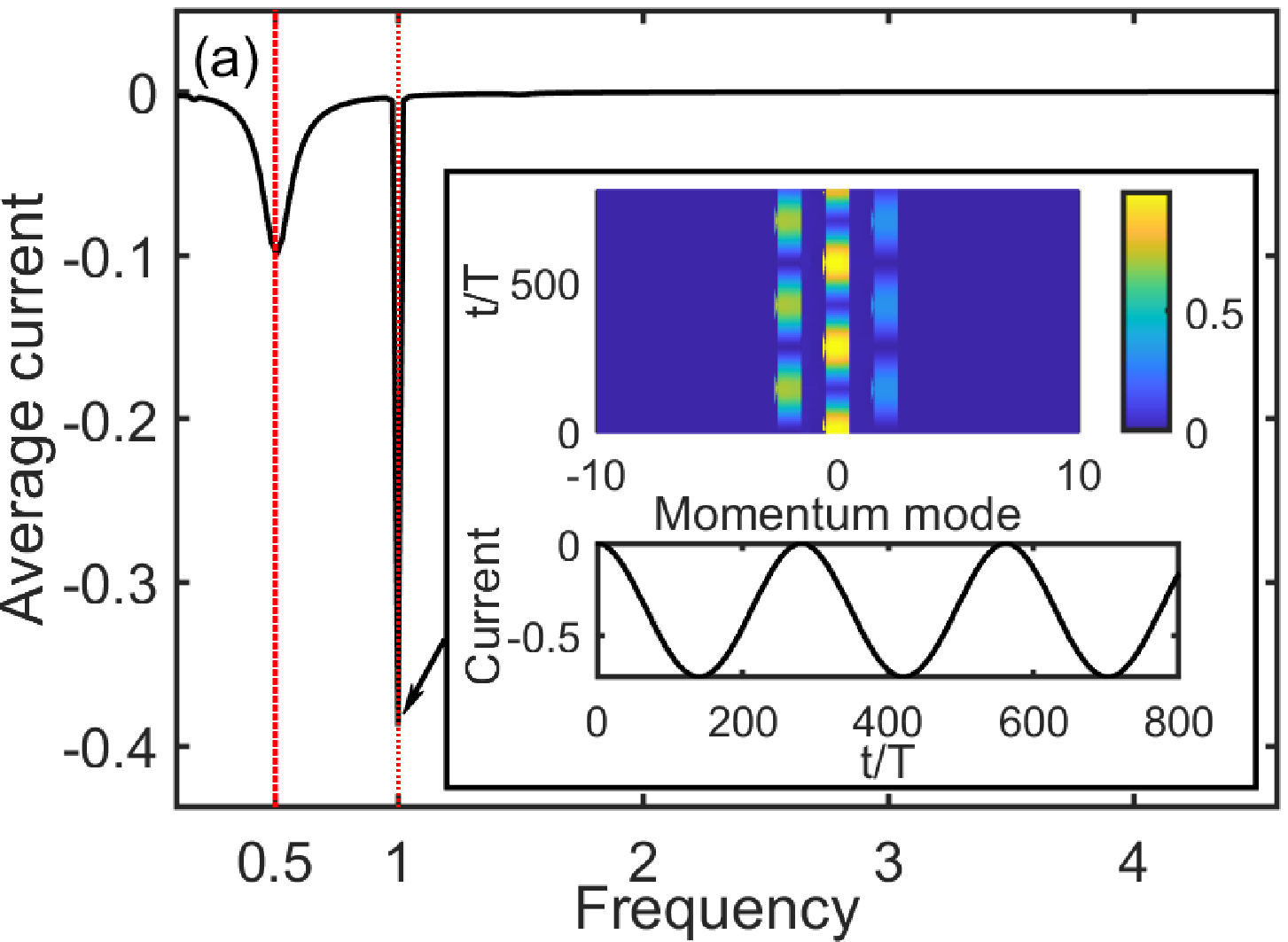}\label{fig3a}}
\subfigure{\centering\includegraphics[scale=0.35]{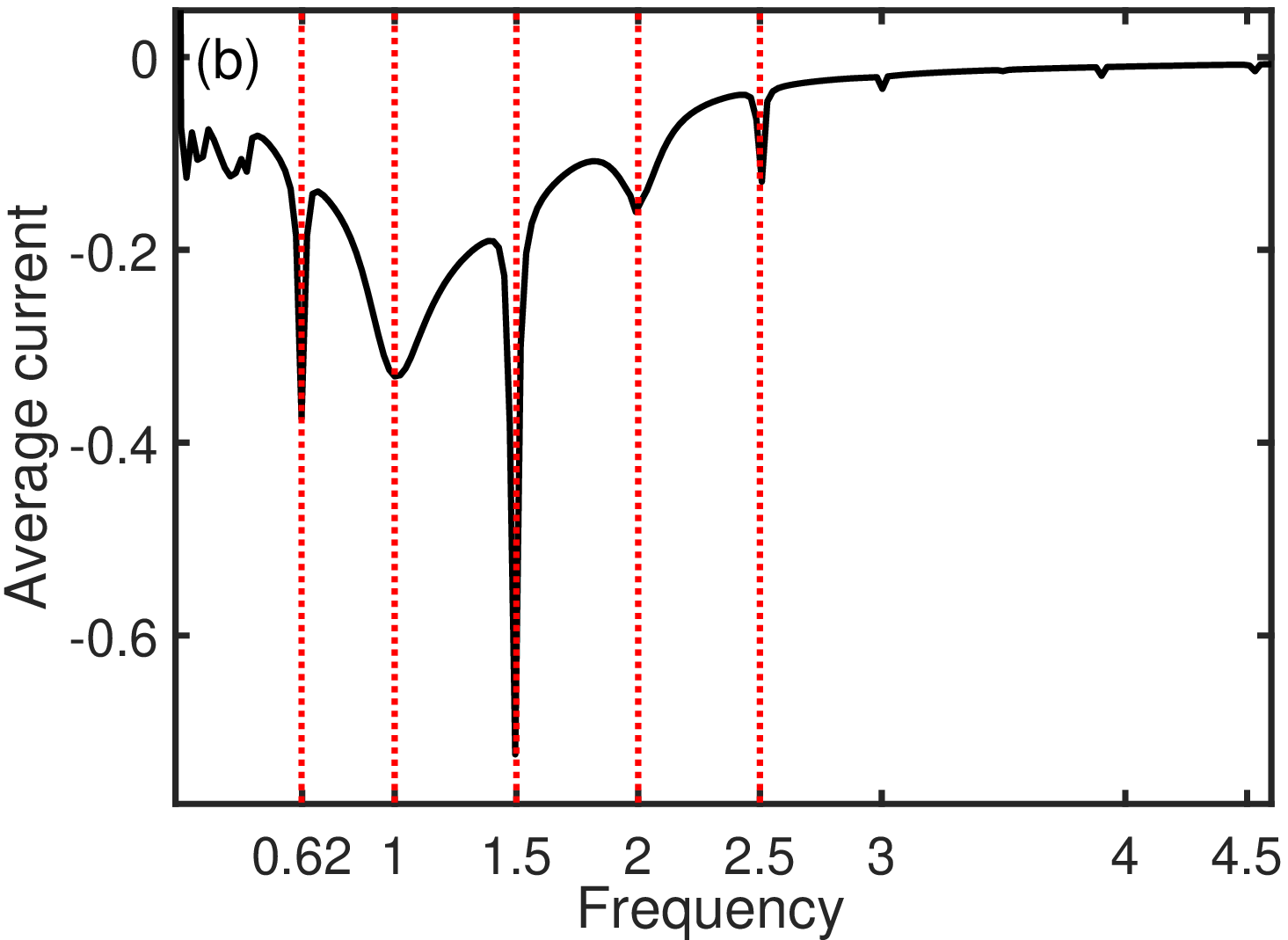}\label{fig3b}}
\subfigure{\centering\includegraphics[scale=0.35]{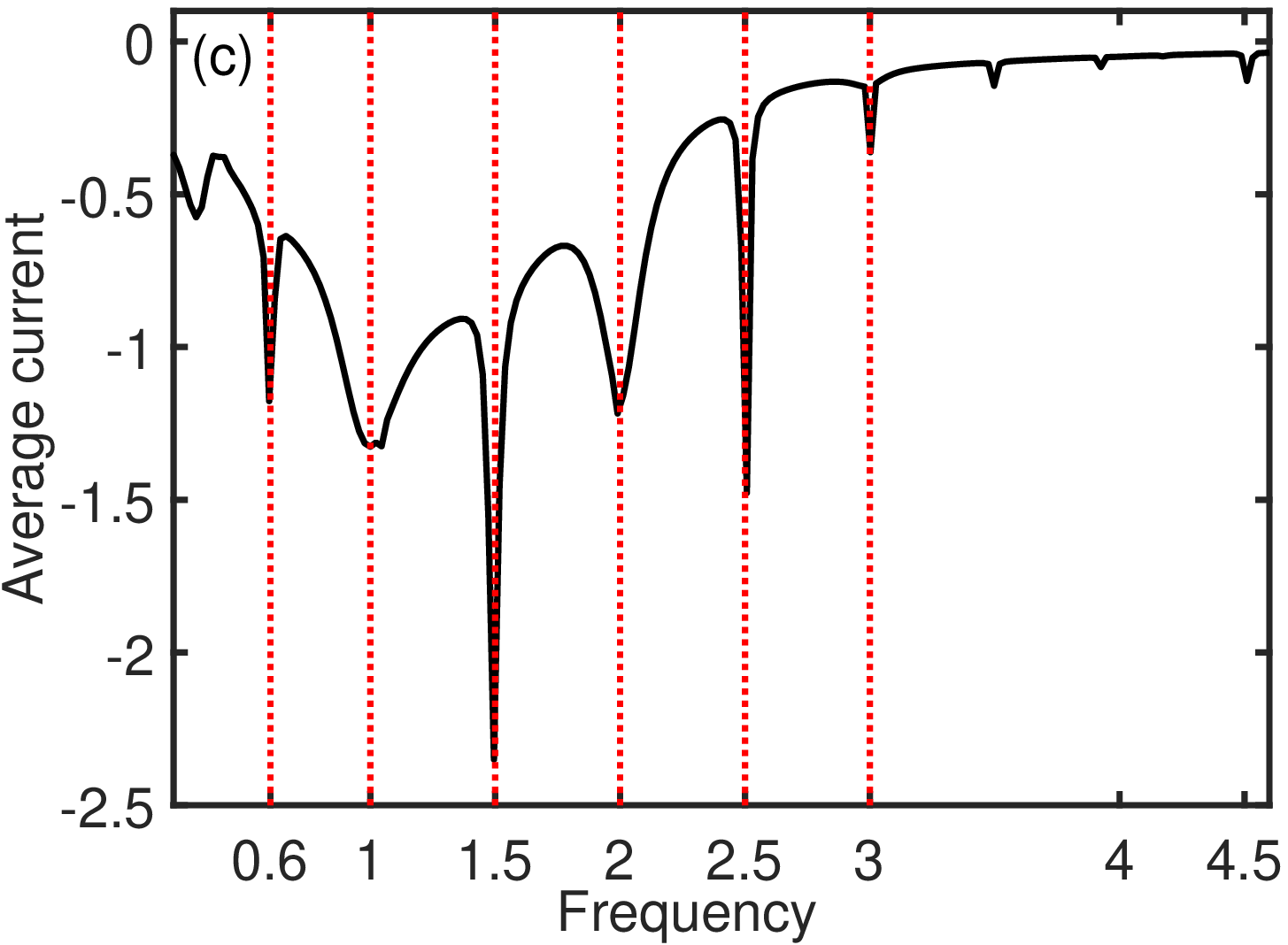}\label{fig3c}}
	\caption{Time-averaged current (TAC) , plotted as a function of the driving frequency $\omega$. (a): TAC for weak driving with $K = 0.1$ and $\uplambda = 0.1$. The TAC peaks only at $\omega = 0.5$ and $\omega = 1$, in agreement with the resonance condition $\frac{n ^2}{2}=m\omega$, representing the first-order and the second-order dynamics, respectively. (b): TAC for strong driving with $K = 1$ and $\uplambda = 0.1$. More resonances are observed for the strong driving and the pronounced resonance is shifted from $\omega=1$ to $\omega=1.5$. (c): The same as in (b), but with a strong non-Hermitian parameter $\uplambda=0.5$. As we can see, the TAC at resonant frequencies is significantly enhanced by increasing the non-Hermitian strength $\uplambda$. Insets in (a) are for the detailed current behaviour at $\omega=1$ resonance: top: time evolution of the populations in momentum space; bottom: Rabi oscillation of the current $I(t)$.
		}\label{fig3}
\end{figure*}

In this section we are particularly interested in a special type of transport phenomenon in the Floquet $\mathcal{PT}$-symmetric system when the conditions for quantum resonance are fulfilled, i.e., $\frac{n^2}{2}=m\omega$, where $\frac{n^2}{2}$ is the  unperturbed levels difference between $\ket{0}$ and $\ket{n}$. For this purpose, we introduce the asymptotic time-averaged current (TAC), which is defined as
\begin{equation}\label{eq6}
	\bar{I}=\lim_{t\to \infty}\frac{1}{t}\int_{t_0=0}^tdt^{\prime}I(t^\prime),
\end{equation}
where the current $I(t)$ is given by
\begin{equation}\label{eq7}
	I(t)=\frac{\int_0^{2\pi}dx\psi^*(x,t)\hat{p}\psi(x,t)}{\mathcal{N}(t)},
\end{equation}
with $\mathcal{N}(t)=\int_0^{2\pi}dx|\psi(x,t)|^2$ being the norm of the quantum state \cite{Stefano Longhi}. In Fig. \ref{fig3}, we numerically study the TAC over a wide range of driving frequencies with fixed driving strength, focusing on the case of unbroken $\mathcal{PT}$ phase. Fig. \ref{fig3a} shows clear signatures of  resonant currents  at $\omega=0.5$, $1$, for weak driving $K=0.1$,  where the pronounced resonance at $\omega=1$ arises from  the mixing between $\ket{0}$ (the initial state ) and the current-carrying mode $\ket{\pm2}$ (see the insets). For strong driving $K=1$, we find that more resonant currents are visible and the position of the pronounced resonance is shifted to $\omega=1.5$, as shown in Fig. \ref{fig3b}. This is obvious because strong driving will excite a larger number of higher momentum modes. Meanwhile, if we keep $K=1$ unchanged and increase the non-Hermitian strength $\uplambda$, we observe that the resonant currents are preserved and that the position of the pronounced resonance does not shift with increasing non-Hermitian strength, as can be seen in Fig. \ref{fig3c}. More interestingly, it can be seen from the comparison of Fig. \ref{fig3b} and \ref{fig3c} that non-Hermiticity enhances the resonant currents.

The generation of resonant currents in the $\mathcal{PT}$-symmetric system is essentially due to the time-reversal symmetry breaking induced by the introduction of non-Hermiticity. This is in contrast to the scheme for achieving directed current in Hermitian systems, where a rachet potential with asymmetric temporal driving is required to simultaneously break the time-reversal and space-inversion symmetries \cite{DenisovS,M. Heimsoth}. In what follows, we attempt to understand such non-Hermitian resonant currents by means of the well-established Floquet theory. In the basis of the Floquet eigenstates, an arbitrary time-evolved quantum state can be expanded as
 \begin{equation}\label{eq8}
 	\ket{\psi(t)}=\sum_\alpha C_\alpha e^{-i\varepsilon_\alpha t}\ket{\phi_\alpha(t)},
 \end{equation}
where $C_\alpha$ is the projection coefficient of the initial state onto the Floquet state $\ket{\phi_\alpha(0)}$, such that $C_\alpha=\braket{\phi_\alpha(0)|\psi(0)}$. Thus, in the unbroken $\mathcal{PT}$ phase, the momentum expectation value is given by
 \begin{align}\label{eq9}
	\braket{p(t)}=&\nonumber\sum_{\alpha^\prime\alpha}{C^*_{\alpha^\prime}}C_\alpha e^{i(\varepsilon_{\alpha^\prime}-\varepsilon_{\alpha})t}\bra{\phi_{\alpha^\prime}(t)}\hat{p}\ket{\phi_{\alpha}(t)}\\\nonumber
	=&\sum_\alpha|C_\alpha|^2\braket{\phi_\alpha(t)|\hat{p}|\phi_\alpha(t)}\\+&\sum_{\alpha^\prime\ne\alpha}{C^*_{\alpha^\prime}}C_\alpha e^{i(\varepsilon_{\alpha^\prime}-\varepsilon_{\alpha})t}\bra{\phi_{\alpha^\prime}(t)}\hat{p}\ket{\phi_{\alpha}(t)},
\end{align}
where $\braket{\phi_\alpha(t)|\hat{p}|\phi_\alpha(t)}$ is the mean instantaneous momentum of the Floquet state, which is a periodic function due to $\ket{\phi_\alpha(t)}=\ket{\phi_\alpha(t+T)}$. In cases other than exact degeneracies, the off-diagonal interference terms $\exp{[i(\varepsilon_{\alpha^\prime}-\varepsilon_{\alpha})t]}$ would average to zero for long enough times, and the contributions to the directed current are given only by the diagonal terms $\Big<\sum_\alpha|C_\alpha|^2\braket{\phi_\alpha|\hat{p}|\phi_\alpha}\Big>_T$, where $\Big<...\Big>_T$ denotes the time average over the  period $T$. Since the non-Hermitian driving potential breaks time-reversal symmetry, there exist asymmetric Floquet states in the unbroken phase, which can be either degenerate or non-degenerate, as we have shown in Fig. \ref{Fig2}. Despite the existence of the degenerate Floquet states, which are well localized in momentum space, they do not overlap with the initial state $\ket{0}$, i.e. the corresponding expansion coefficients $C_\alpha$ are zero, thus having no consequence on the currents. In our model, the numerical results show that all the asymmetric non-degenerate Floquet states overlap with the initial state $\ket{0}$ and have a negative mean momentum, which contributes to the diagonal terms ($\alpha=\alpha^\prime$) in Eq. (\ref{eq9}), thus leading to the appearance of directed currents as shown in Fig. \ref{fig3}.

As shown in Fig. \ref{fig3}, the non-Hermitian ratchet current can be significantly increased for a finite set of resonant driving frequencies. To give an excellent account of the resonant currents, we follow the perturbative method of \cite{M. Heimsoth} and generalize it to the non-Hermitian model, which is found to be in surprisingly good agreement with numerical results. Let us focus on the case $\omega=1$, which specifies the resonance condition $\frac{n^2}{2}=m\omega$ with $m=2$ and $n=\pm2$, as an example to illustrate this perturbative analysis method. Our starting point for the study is given by the unperturbed Floquet states $\ket{n,m}=\frac{1}{\sqrt{2\pi T}}e^{inx}e^{-im\omega t}$, which are
Floquet eigenstates of the unperturbed Floquet Hamiltonian $\frac{\hat{p}^2}{2}-i\partial/\partial t$, with the Floquet quasienergies $\varepsilon=\frac{n^2}{2}-m\omega$. We expect to find the highest values of the directed current when the resonance condition $\frac{n^2}{2}=m\omega$ is fulfilled, i.e. the static energy difference is bridged by the energy of an integer number of photons. In such cases, all these unperturbed  Floquet states satisfying the resonance condition are degenerate with $\varepsilon=0$ and connected by the periodic driving composed of a finite number of Fourier components.  When the initial state is prepared as the ground state of the unperturbed Hamiltonian, which corresponds to the Floquet state $\ket{0,0}$, we may expect the driving
to mix $\ket{0,0}$ with the other unperturbed Floquet states $\ket{n,m}$ satisfying the resonance condition $m = \frac{n^2}{2\omega}$. For example, for the case $\omega=1$, the system is expected to evolve from $\ket{0,0}$ towards $\ket{2,2}$ and $\ket{\bar{2},2}$ (for notational brevity, we use  $\bar{n} \equiv -n$), where the mixing with the higher-lying ($n > 2$) resonant Floquet states is very small and can be neglected under weak driving. Thus, when $\omega=1$, truncating the Hilbert space to just these three resonant states $\{\ket{2,2},\ket{0,0},\ket{\bar{2},2}\}$ is sufficient to describe the dynamics under weak driving.  Of course, such
a truncation can be applied to other resonant frequencies (e.g. $\omega=0.5$) as
well, where the resonant Floquet states $\ket{n,m}$ involved are different for different driving frequencies.  By applying the time-independent perturbation theory in the extended Hilbert space spanned by the relevant three resonant Floquet states $\{\ket{2,2},\ket{0,0},\ket{\bar{2},2}\}$, the dynamics of the system can be described by an effective three-level non-Hermitian model for $\omega=1$ (more details can be seen in the Appendix)
 \begin{equation}\label{eq10}
 	T\simeq\left[\begin{array}{ccc}
 	0 & \Gamma_{-} & 0 \\
 	\Gamma_{+} & 0 & \Gamma_{-} \\
 	0 & \Gamma_{+} & 0
 \end{array}\right],
 \end{equation}
where $\Gamma_{\pm}=K^2\uplambda_\pm^2/8$, and $\uplambda_\pm\equiv \uplambda\pm1$. We can see that the effective $T$-matrix is non-Hermitian due to $\uplambda\ne0$. By solving the Schr{\"o}dinger equation on the basis of the effective three-level model, we obtain the time evolution of the populations on the momentum eigenstates as follows,
	\begin{align}\label{eq11}
		&P_2(t)=\frac{\Gamma_-^2}{\Omega^2}\sin^2(\Omega t)\nonumber\\
		&P_0(t)=\cos^2(\Omega t)\\
		&P_{-2}(t)=\frac{\Gamma_+^2}{\Omega^2}\sin^2(\Omega t)\nonumber,
	\end{align}
where $P_n$ denotes the population in the momentum mode $\ket{n}$, and the Rabi oscillation frequency is given by $\Omega=\sqrt{2\Gamma_{-}\Gamma_{+}}$. The corresponding norm thus reads  $\mathcal{N}(t)=P_{-2}+P_0+P_2=\cos^2\Omega t+\frac{\Gamma_-^2+\Gamma_+^2}{\Omega^2}\sin^2\Omega t$. By definition, Eq. (\ref{eq7}), after omitting the contribution of the norm to the current behavior, the current is given by
 \begin{equation}\label{eq12}
	I(t)=\frac{-2(\Gamma_+^2-\Gamma_-^2)}{(\Gamma_+^2+\Gamma_-^2)+\Omega^2\cot^2(\Omega t)}.
\end{equation}
When the non-Hermitian strength $\uplambda=0$ is zero, we have $\Gamma_+=\Gamma_-$ such that $I(t) = 0$, which is a natural consequence due to the fact that the limit of $\uplambda \to 0$ leads to the recovery of the time-reversal symmetry. In this way, we analytically confirm the conclusion that the non-Hermiticity produces the resonant current in the system with symmetric temporal driving, which would otherwise forbid the  generation of direct currents.

Considering the other resonant frequency $\omega = 0.5$, the weakly driven system can also be truncated to an effective three-level model operating in a reduced Hilbert space of resonant Floquet states  $\ket{n,m}=\{\ket{1,1},\ket{0,0},\ket{\bar{1},1}\}$. The corresponding effective $T$-matrix  is given by
 \begin{equation}\label{eq13}
	T\simeq\left[\begin{array}{ccc}
		0 & -\Gamma^\prime_{-} & 0 \\
		-\Gamma^\prime_{+} & 0 & \Gamma^\prime_{-} \\
		0 & \Gamma^\prime_{+} & 0
	\end{array}\right],
\end{equation}
where $\Gamma^\prime_\pm=K(1\pm\uplambda)/4$. A comparison of the $T$-matrices at $\omega = 0.5$ and $\omega = 1$ shows that the effective coupling $\Gamma_ \pm$ is proportional to $K^2$ when $\omega = 1$, while when $\omega = 0.5$, the effective coupling $\Gamma^\prime_\pm$ is proportional to $K$. This is because for $\omega = 1$ the mixing between $\ket{0,0}$ and the states $\ket{2,2}$ and $\ket{\bar{2},2}$ is the second-order transition via the virtual intermediate states $\ket{1,1}$ and $\ket{\bar{1},1}$, whereas for $\omega=0.5$ the mixing between $\ket{0,0}$ and the states $\ket{1,1}$ and $\ket{\bar{1},1}$ is the direct, first-order transition. Following the same line of reasoning as above, we can obtain the populations in the momentum modes $\ket{0}$ and $\ket{\pm 1}$, from which we can further obtain the norm $\mathcal{N}(t)=\frac{\Gamma^{\prime2}_++\Gamma^{\prime2}_-}{|\Omega^\prime|^2}|\sin(\Omega^\prime t)|^2+|\cos(\Omega^\prime t)|^2$, and the current as
\begin{equation}\label{eq14}
	I^\prime(t)=-\frac{\Gamma^{\prime2}_+-\Gamma^{\prime2}_-}{(\Gamma^{\prime2}_++\Gamma^{\prime2}_-)+|\Omega^{\prime2}||\cot(\Omega^\prime t)|^2},
\end{equation}
in which $\Omega^{\prime}=\sqrt{2\Gamma^\prime_{-}\Gamma^\prime_{+}}$.

Let us further examine the second-order dynamics for the $\omega=1$ resonance and the first-order dynamics for the $\omega=0.5$ resonance. For the former, the Rabi oscillation frequency $\Omega=\sqrt{2\Gamma_{-}\Gamma_{+}}$ is always real, while for the latter, $\Omega^{\prime}=\sqrt{2\Gamma^\prime_{-}\Gamma^\prime_{+}}$ becomes purely imaginary when $\uplambda > 1$. As such, the analytical current Eq. (\ref{eq12}) for $\omega=1$ resonance marks a periodic oscillation, implying that the second-order process can only describe the unbroken phase dynamics. On the other hand, for the $\omega=0.5$ resonance, as $\uplambda$ is increased beyond the parameter point $\uplambda=1$, the periodic oscillation of the current [see Eq. (\ref{eq14})] disappears and is replaced by a hyperbolic behavior, with the norm growing exponentially and the current approaching an asymptotic value of $-1$. This implies that $\uplambda=1$ signals the onset of a $\mathcal{PT}$ symmetry phase transition for the $\omega=0.5$ resonance, which agrees exactly with the numerical results shown in Sec. \ref{II} for weak driving, and at the same time indicates that the perturbative analytical formula for the first-order process is also applicable in the broken phase.

\begin{figure}[htp]
	\center
	\includegraphics[width=7cm]{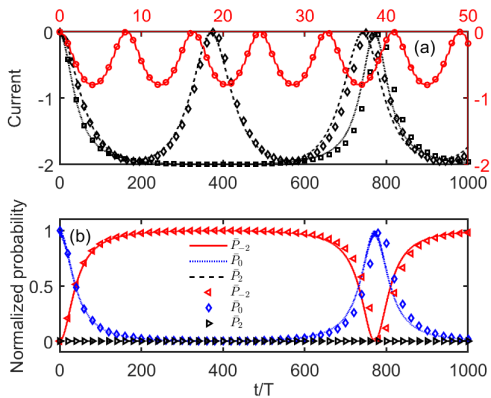}
	\caption{Comparison of the exact numerical results (shown by lines) with the analytical results predicted by the effective three-level non-Hermitian model (circles, triangles, diamonds and squares)  in the $\mathcal{PT}$-unbroken regime with weak driving $K = 0.1$. (a): Time evolution of the current $I(t)$, with different parameters $\omega = 0.5$, $\uplambda = 0.5$ (solid line and circles), $\omega = 1$, $\uplambda = 0.5$ (dashed line and diamonds), and $\omega = 1,$ $\uplambda = 0.8$ (dotted line and squares). (b): Time-dependent normalized probabilities $\bar{P}_n(t)=P_n(t)/\mathcal{N}(t)$, corresponding to the current shown by the dotted line and squares with fixed parameters $\omega=1, \uplambda=0.8$.}
	\label{fig4}
\end{figure}

Fig. \ref{fig4} shows an extremely good agreement between the numerical results and the analytical results of the time-dependent currents and the normalized probabilities for a weak driving $K = 0.1$. Since the effective coupling $\Gamma_{\pm}\sim K^2$ for the $\omega=1$ resonance is smaller than the one ($\Gamma^\prime_{\pm} \sim K$) for the $\omega=0.5$ resonance, the former oscillation period is an order of magnitude larger than the latter, and for better comparison we use different axis scales (red for $\omega=0.5$ and black for $\omega=1$) to show the current behaviors. As shown in Fig. \ref{fig4} (a), the current for the $\omega = 0.5$ resonance shows the Rabi-like oscillation, but for the $\omega = 1$ resonance, the currents show the deviation from the Rabi-like oscillation. When $\uplambda$ increases, the departure from Rabi-like oscillation becomes drastic, deforming into a square-wave-like shape with a significant duration of staying at a steady value of $-2$ for the current with $\omega=1$ resonance. This process is accompanied by the significant duration of the normalized probability staying at $\bar{P}_{-2}(t) \approx  1$, $\bar{P}_2(t) \approx \bar{P}_0(t) \approx  0$, as shown in Fig. \ref{fig4} (b), implying that TAC will be enhanced upon increasing the non-Hermitian strength.


\begin{figure}[htp]	
	\center
	\includegraphics[width=7cm]{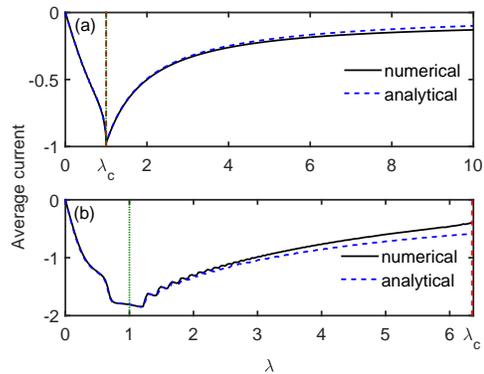}
	\caption{Time-averaged current versus non-Hermitian parameter $\uplambda$ with weak driving $K = 0.1$, for (a): the $\omega = 0.5$ resonance; (b): the $\omega = 1$ resonance. The green dotted vertical lines mark the $\uplambda = 1$, and the  red dashed vertical lines mark the spontaneous $\mathcal{PT}$ symmetry breaking threshold  $\uplambda_c$. We can see that for weak driving strength the agreement between the analytical results predicted by the three-level model and the exact numerical results is excellent, even in the $\mathcal{PT}$-broken regime ($\uplambda>1$) for the $\omega=0.5$ resonance.}
	\label{fig5}
\end{figure}

In addition, in Fig. \ref{fig5}, we numerically investigate the TAC as a function of the non-Hermitian strength $\uplambda$ for both the $\omega=0.5$ resonance [Fig. \ref{fig5} (a)] and the $\omega=1$ resonance [Fig. \ref{fig5} (b)], which show good agreement with the analytical results derived from the effective three-level model. In both cases, when $\uplambda < 1$, the time-averaged currents are enhanced with increasing non-Hermitian strength $\uplambda$, while above the parameter point $\uplambda = 1$ (marked by the green vertical line), the time-averaged currents are instead reduced with increasing non-Hermitian strength $\uplambda $. The boundary $\uplambda=1$, separating the two regions with physically different dependence of the current on non-Hermitian strength, corresponds exactly to  the symmetry breaking point for the $\omega=0.5$ resonance, but is located deeply in the $\mathcal{PT}$-unbroken regime for the $\omega=1$ resonance (the symmetry breaking threshold is marked by $\uplambda_c$). What's more, for the $\omega = 0.5$ resonance, the analytical result of the three-level model is still applicable in the $\mathcal{PT}$-broken regime ($\uplambda>1$), showing a good agreement with the direct numerical result based on Eq. (\ref{fig1}).

\section{Directed currents at exceptional points}\label{IV}

An interesting property of non-Hermitian systems is the existence of spectral singularities, known as exceptional points (EP)\cite{A A Zyablovsky,W D Heiss}, which has led to a variety of novel phenomena and fascinating applications. At this point, not only are the eigenvalues identical, but also the corresponding eigenvectors coalesce to one. For $\mathcal{PT}$-symmetric systems, it turns out that the spontaneous symmetry breaking points, beyond which the spectrum changes from all real to complex, are just those values at which an EP of the system appears.

\begin{figure}[htp]
	\centering
	\includegraphics[width=7cm]{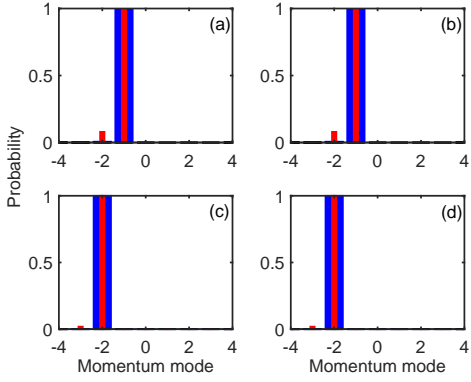}
	\caption{Probabilities (blue thick bar) and real parts of the probability amplitudes (red thin bar)  in different momentum modes for  two pairs  of coincident Floquet states, corresponding to quasienergy -0.498 [upper panels (a) and  (b)]  and 0.007 [lower panels (c) and (d)]. We see that the two quasienergies are identical, and the two Floquet eigenvectors are also completely identical, which is indeed the characteristic of exceptional point (EP) behaviour. The parameters are $K=0.1, \omega=1, \uplambda=1$.}
	\label{fig6}
\end{figure}

\begin{figure}[htp]
	\centering
	\includegraphics[width=7cm]{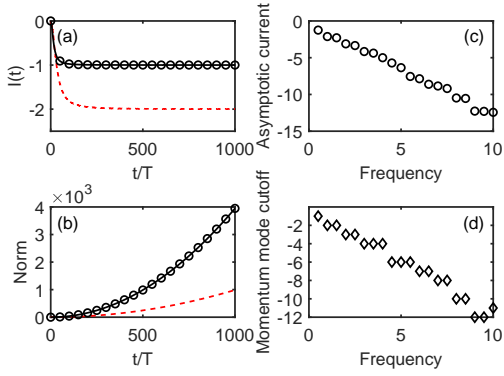}
	\caption{ (a): Time dependence of current $I(t)$ with different parameters $K = 0.01$, $\omega = 0.5$ (solid line) and $K = 0.1$, $\omega = 1$ (dashed line) at EPs. (b): Power-law growth of the norm at EPs, for $K = 0.01$, $\omega = 0.5$ (solid line), and $K = 0.1$, $\omega = 1$ (dashed line). The analytical results (circles) for  the norm,  predicted by Eq. (\ref{eq19}), and for the current, predicted by Eq. (\ref{eq20}), agree well with the numerical results.
	(c): Asymptotic current versus frequency $\omega$ at EPs. (d): The momentum mode cut-off (i.e. the maximum attainable negative momentum mode) versus the driving frequency $\omega$ at EPs. Here, the EPs correspond to  the parameter point $\uplambda = 1$ and $\omega = 0.5l$ with integer $l$.}
	\label{fig7}
\end{figure}

The situation is richer when a non-Hermitian system is periodically driven. As discussed above, for the $\omega = 0.5$ resonance, the parameter point $\uplambda = 1$ corresponds to the symmetry breaking point and, consequently, to the EP. Our further numerical investigation shows that for general $\omega = 0.5l$ with integer $l$, the parameter point $\uplambda = 1$ (the real and imaginary parts of the potential depth are equal) is very unique, where the current and the Floquet spectrum behave unexpectedly, regardless of whether $\uplambda = 1$ is the symmetry breaking point. If $l$ is odd, $\uplambda = 1$ represents the spontaneous $\mathcal{PT}$ symmetry breaking point. However, if $l$ is even, $\uplambda = 1$ is below the phase transition point, but it is an EP where the quasienergies and the Floquet states coincide simultaneously. As shown in Fig. \ref{fig6}, for $K = 0.1$ and $\omega = 1$, $\uplambda = 1$ (as pointed out in Fig. \ref{fig1}, it is not a symmetry breaking point) indeed represents an EP, where we observe the two pairs of coincident (identical) Floquet states, corresponding to the degenerate quasienergy -0.498 (upper panels) and the degenerate quasienergy 0.007 (lower panels), respectively. Under such a circumstance, there appear a persistent current with an asymptotically constant value and a power-law increase of the norm $\mathcal{N}(t) \propto t^2$ (see Fig. \ref{fig7}), even though $\uplambda = 1$ is not the symmetry breaking point and is deeply rooted in the $\mathcal{PT}$-unbroken regime.

We analytically investigate the current and population behaviors at the exceptional point using a specific driving frequency $\omega=0.5$. The Schr{\"o}dinger equation (\ref{eq1}) in the momentum representation can be written as
\begin{equation}\label{eq15}
	i\frac{\partial}{\partial t}c_n=\frac{k^2}{2}c_n+\frac{iK\sin(\omega t)}{2}(\uplambda_+c_{n+1}+\uplambda_-c_{n-1}),
\end{equation}
where $c_n$ denotes the population amplitude on the momentum mode $\ket{n}$ and the coupling coefficients are given by $\uplambda_+= \bra{n}\hat{V}\ket{n+1}$ and $\uplambda_-= \bra{n}\hat{V}\ket{n-1}$. Since the coupling strength in the hopping direction to the negative momentum state is greater than to the positive momentum state, a negative average current is generally expected. At the exceptional point $\uplambda=1$, Eq. ($\ref{eq15}$) is reduced to
\begin{equation}\label{eq16}
	i\frac{\partial}{\partial t}c_n=\frac{k^2}{2}c_n+\frac{iK\sin(\omega t)}{2}(\uplambda_+c_{n+1}).
\end{equation}
To be specific,  we consider the weakly driven system with the resonance frequency $\omega = 0.5$, starting from the initial state $c_n(0)=\delta_{n,0}$, then the dynamics is limited in the subspace spanned by the basis ${\ket{n}=\ket{-1},\ket{0},\ket{1}}$, and Eq. (\ref{eq16}) is truncated  to
\begin{equation}\label{eq17}
	i\frac{\partial}{\partial t}\begin{bmatrix}
		c_{-1}(t)\\
		c_0(t)\\
		c_1(t)
	\end{bmatrix}=
	\begin{bmatrix}
		\frac{1}{2}& iK\sin(\omega t) &0 \\
		0& 0 & iK\sin(\omega t) \\
		0& 0 &\frac{1}{2}
	\end{bmatrix}
	\begin{bmatrix}
		c_{-1}(t)\\
		c_0(t)\\
		c_1(t)
	\end{bmatrix}.
\end{equation}
The solution of Eq. (\ref{eq17}) can be obtained exactly as follows
\begin{equation}\label{eq18}
\left\{\begin{array}{l}
	c_{-1}(t)=\frac{iK}{2}te^{-\frac{it}{2}}-iK\sin(\omega t) \\
	c_0(t)=1\\
	c_{1}(t)=0,
\end{array}\right.
\end{equation}
which gives the power-law increase of the norm
\begin{equation}\label{eq19}
	\mathcal{N}(t)\simeq 1+K^2t^2/4.
\end{equation}
As time tends to infinity, the asymptotic current reads
\begin{equation}\label{eq20}
	I(t\to \infty)=\lim_{t\to\infty}\frac{-1}{1+\frac{4}{K^2t^2}}=-1.
\end{equation}

Our theoretical prediction is verified by numerical results, as shown in Fig. \ref{fig7} [see  the analytical results (circles) and the numerical results (black solid lines) in Figs. \ref{fig7} (a) and (b) ]. We also numerically investigate the behaviour of the current and the norm for $\omega = 1$, as shown in Figs. \ref{fig7} (a) and (b) (see the red dashed lines), where we observe that the magnitude of the asymptotic current is larger than its counterpart for $\omega = 0.5$.
This is evident from the fact that for $\omega = 1$, the system can be excited to a higher momentum state $\ket{-2}$. The dependence of the asymptotic current on the resonant frequency $\omega$ is also investigated numerically, and it is found  in Fig. \ref{fig7}(c) that the asymptotic
current increases as the resonant frequency $\omega$ increases. Such a dependence stems from the fact
that the momentum mode cut-off (i.e. the maximum attainable negative momentum mode) increases with the resonant frequency, see Fig. \ref{fig7}(d). In our work, our focus is only on the case of $\phi = 0$. If we set the initial phase $\phi = \pi$, the coupling coefficients are reversed, i.e. $\uplambda_+ = \bra{n+1}\hat{V}\ket{n}$ and $\uplambda_- = \bra{n-1}\hat{V}\ket{n}$, leading to a positive mean current, which opens a new avenue to steer the direction of the directed currents.

\section{$\mathcal{PT}$-symmetry-breaking-induced ratchet currents}\label{V}

\begin{figure}[htp]
	\center
	\includegraphics[width=7cm]{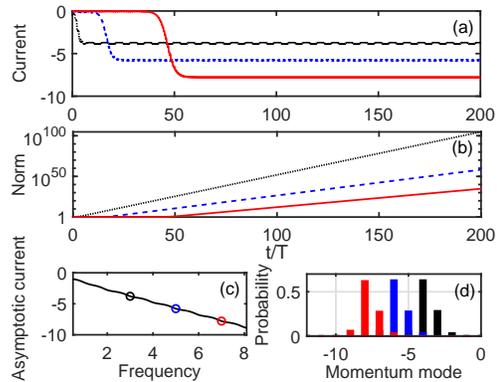}
	\caption{(a): Time evolution of the current in the $\mathcal{PT}$-broken regime at different driving frequencies, $\omega = 3$ (black dotted line), $\omega = 5$ (blue dashed line) and $\omega = 7$ (red solid line).  (b): Time evolution of the corresponding norms. (c): Asymptotic current versus driving frequency $\omega$. The circles in different colors represent the corresponding asymptotic currents [the saturation levels of $I(t)$ in panel (a)]. (d): The Floquet eigenstates in the momentum space with the largest imaginary quasienergy for $\omega = 3$ (black), $\omega = 5$ (blue) and $\omega = 7$ (red), which give rise to the three different asymptotic currents in panel (c) marked by circles,  respectively. In all plots the parameters are $K = 1$, $\uplambda = 1.5$.}
	\label{fig8}
\end{figure}

In this section, we turn our attention to the current dynamics of the $\mathcal{PT}$-broken regime, in which the norm starts to grow exponentially. Fig. \ref{fig8} (a) shows the numerical results for the time evolution of the  current  in the $\mathcal{PT}$-broken regime at different driving frequencies. As we can see, the current evolves to a non-zero asymptotic value for all driving frequencies and the asymptotic current grows as the driving frequency increases, which bears a close resemblance to the current behaviour at EPs as shown in Fig. \ref{fig7} (a). The main difference is that for a fixed non-Hermitian parameter, the asymptotic current in the $\mathcal{PT}$-broken regime can be generated for all continuously-varying driving frequencies, whereas the asymptotic current at EPs can only be generated for certain discrete driving frequencies. The time evolution of the corresponding norms is illustrated in Fig. \ref{fig8} (b), where we observe that the norm grows exponentially and accelerates as the driving frequency increases. The enhancement of the asymptotic current by increasing the driving frequency is more clearly demonstrated in Fig. \ref{fig8} (c). In Fig. \ref{fig8} (c), we see that the value of asymptotic current increases almost linearly with the driving frequency.	

The mechanism of the generation of the asymptotic current in the $\mathcal{PT}$-broken regime can be understood as follows. At the initial time, an arbitrary state can be expanded on the basis of the Floquet eigenstates, namely,

\begin{equation}\label{eq21}
	\ket{\psi(0)}=\sum_\alpha C_\alpha\ket{\phi_\alpha(0)},
\end{equation}
where $C_\alpha$ are the amplitudes for the Floquet eigenstates $\ket{\phi_\alpha(0)}$. With the time evolution, the quantum state takes the form of
$
	\ket{\psi(t)}=\sum_\alpha C_\alpha e^{-i\varepsilon_\alpha t}\ket{\phi_\alpha(t)}.
$
In the $\mathcal{PT}$-broken regime, the quasienergy is complex, i.e., $\varepsilon_\alpha=\varepsilon_\alpha^r+i\varepsilon_\alpha^i$. Accordingly, we have
\begin{equation}
	\ket{\psi(t)}=\sum_\alpha C_\alpha e^{-i\varepsilon_\alpha^r t+\varepsilon_\alpha^it}\ket{\phi_\alpha(t)}.
\end{equation}
As time increases, the component $C_\alpha \exp(\varepsilon_\alpha^it)$ with $\varepsilon^i_\alpha > 0$ grows exponentially and that with $\varepsilon^i_\alpha < 0$ decays, so that the quantum state $\ket{\psi(t)}$ eventually evolves to the Floquet eigenstate with the maximum $\varepsilon^i_\alpha$. Thus, the mean momentum $\braket{p}$ corresponding to these Floquet eigenstates with  maximum $\varepsilon^i_\alpha$ is the main contributor to the current. To confirm our analysis, in Fig. \ref{fig8} (d) we present numerically the momentum distributions of the Floquet eigenstates with maximum $\varepsilon^i_\alpha$ for three different driving frequencies (marked by different colors). A detailed inspection shows that the mean momentum $\braket{p}$ corresponding to these Floquet eigenstates increases as the driving frequency increases, and shifts to larger negative values, with each mean momentum having a one-to-one correspondence with the asymptotic current as shown in Fig. \ref{fig8} (a).

\begin{figure}[htp]
	\center
	\includegraphics[width=7cm]{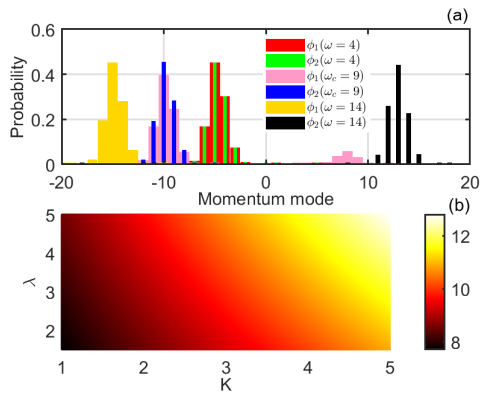}
	\caption{(a): Momentum distributions of the three pairs of degenerate Floquet states with the largest imaginary value of the quasienergy, for $\omega = 4 < \omega$ (red thick and green thin bars), $\omega_c = 9$ (pink thick and blue thin bars), and $\omega = 15 > \omega_c$ (yellow thick and black thin bars). The other parameters are $K = \uplambda = 2$ which corresponds to a broken $\mathcal{PT}$ phase. (b): Threshold frequency $\omega_c$, corresponding to the appearance of the separation of the momentum distribution of the degenerate Floquet eigenstates with the largest imaginary part of the quasienergy, as a function of the driving amplitude $K$ and the non-Hermitian strength $\uplambda$. The different map colors specify different values of $\omega_c$.} \label{fig9}
\end{figure}

Same as in the $\mathcal{PT}$-unbroken regime, we find that most of the Floquet eigenstates in the $\mathcal{PT}$-broken regime are degenerate and well localized in the momentum space,  which results in that Floquet eigenstates with the largest $\varepsilon^i_\alpha$ always appear in pairs as shown in Fig. \ref{fig9} (a). Interestingly, we find that when the driving frequency is below a certain threshold $\omega_c$, these two Floquet eigenstates occupy the same momentum modes (that is, the population probabilities in the momentum modes are identical) with a non-zero and negative mean momentum. This leads to a stable negative asymptotic current.  When the driving frequency is above the threshold, i.e. $\omega > \omega_c$, these two degenerate Floquet eigenstates begin to gradually separate from each other. As $\omega$ increases, these two Floquet eigenstates become more and more separated and populate the momentum modes in an almost symmetrical manner: one with a positive mean momentum and the other with a negative mean momentum. Due to the linear superposition of these two different degenerate eigenstates with the largest $\varepsilon^i_\alpha$, the initial state evolves into one of the two degenerate eigenstates with unpredictability. This means that the system will randomly settle into one of the two different degenerate eigenstates with the largest $\varepsilon^i_\alpha$, leading to an unpredictability of the current direction. Thus, to achieve a stable asymptotic current, we should limit the driving frequency to less than the threshold value, i. e., $\omega < \omega_c$. The phase diagram of the frequency threshold $\omega_c$ in the $(K, \uplambda)$ space is plotted in Fig. \ref{fig9} (b). It can be observed that the threshold value increases with both the driving amplitude $K$ and the non-Hermitian parameter $\uplambda$.

\section{Nonlinearity effects on the non-Hermitian currents}\label{VI}
\begin{figure*}[tbp]
	\subfigure{
		\centering
		\includegraphics[scale=0.35]{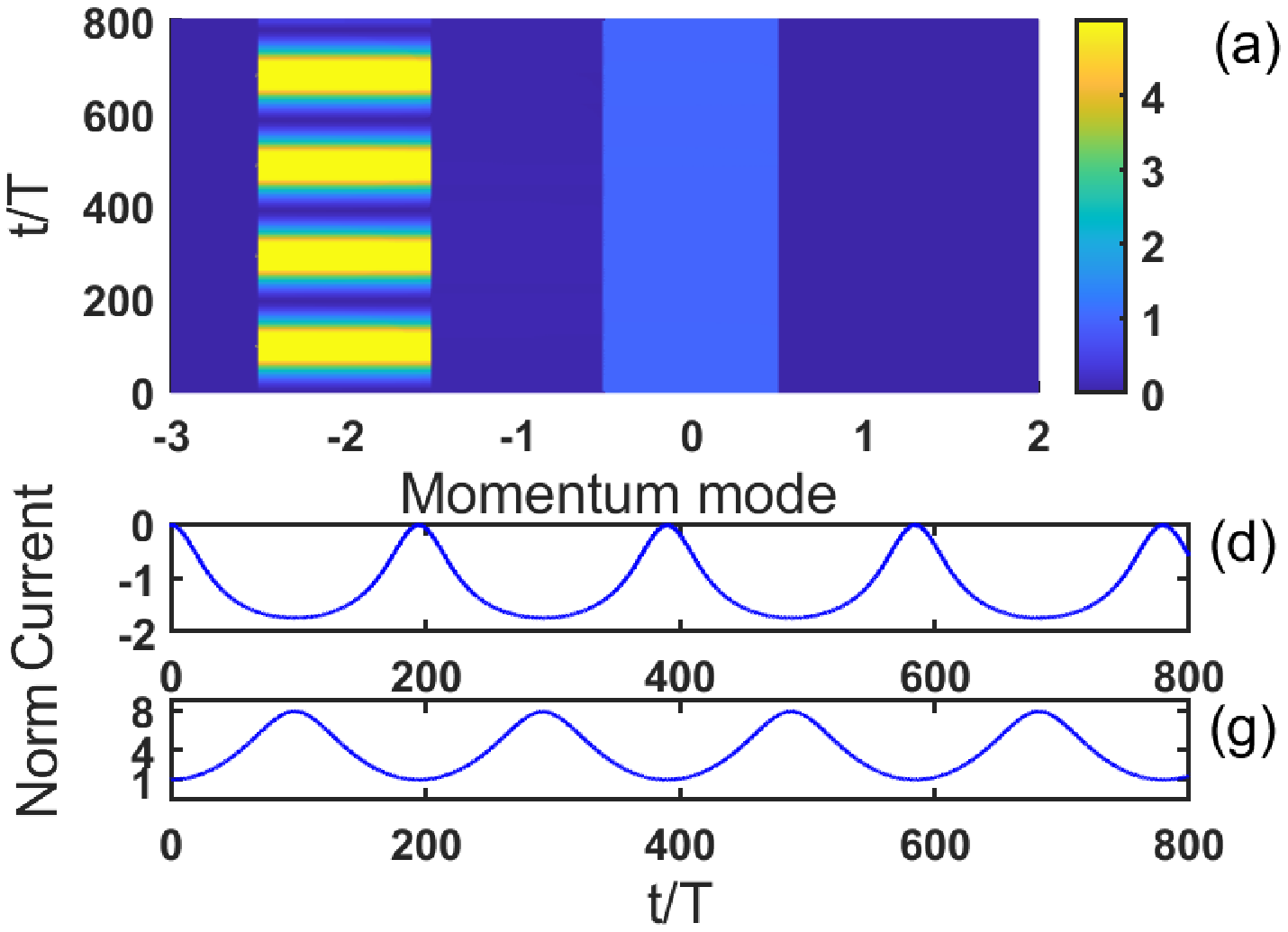}
		\label{fig10a}
	}
	\subfigure{
		\centering
		\includegraphics[scale=0.35]{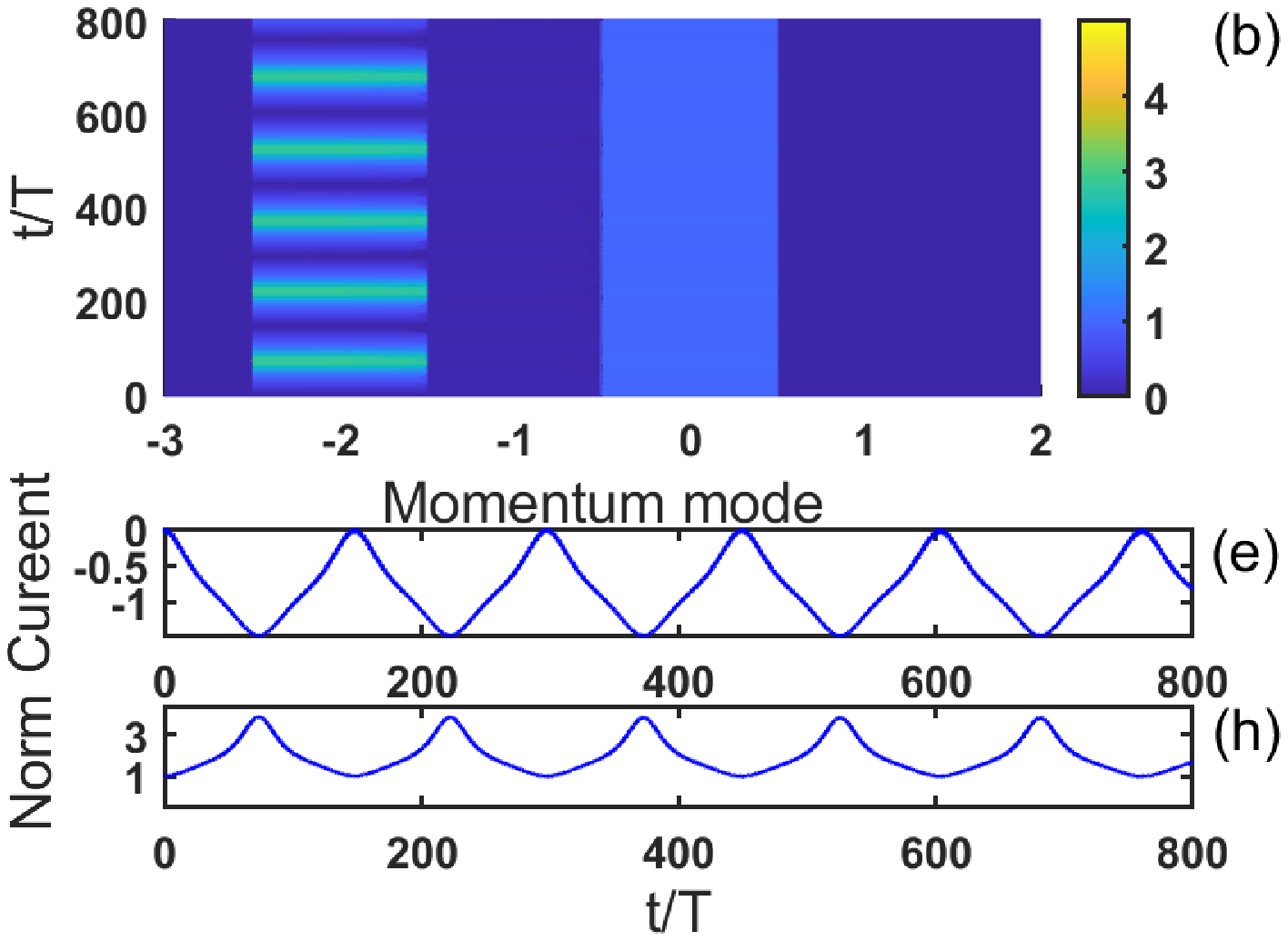}
		\label{fig10b}
	}
	\subfigure{
		\centering
		\includegraphics[scale=0.35]{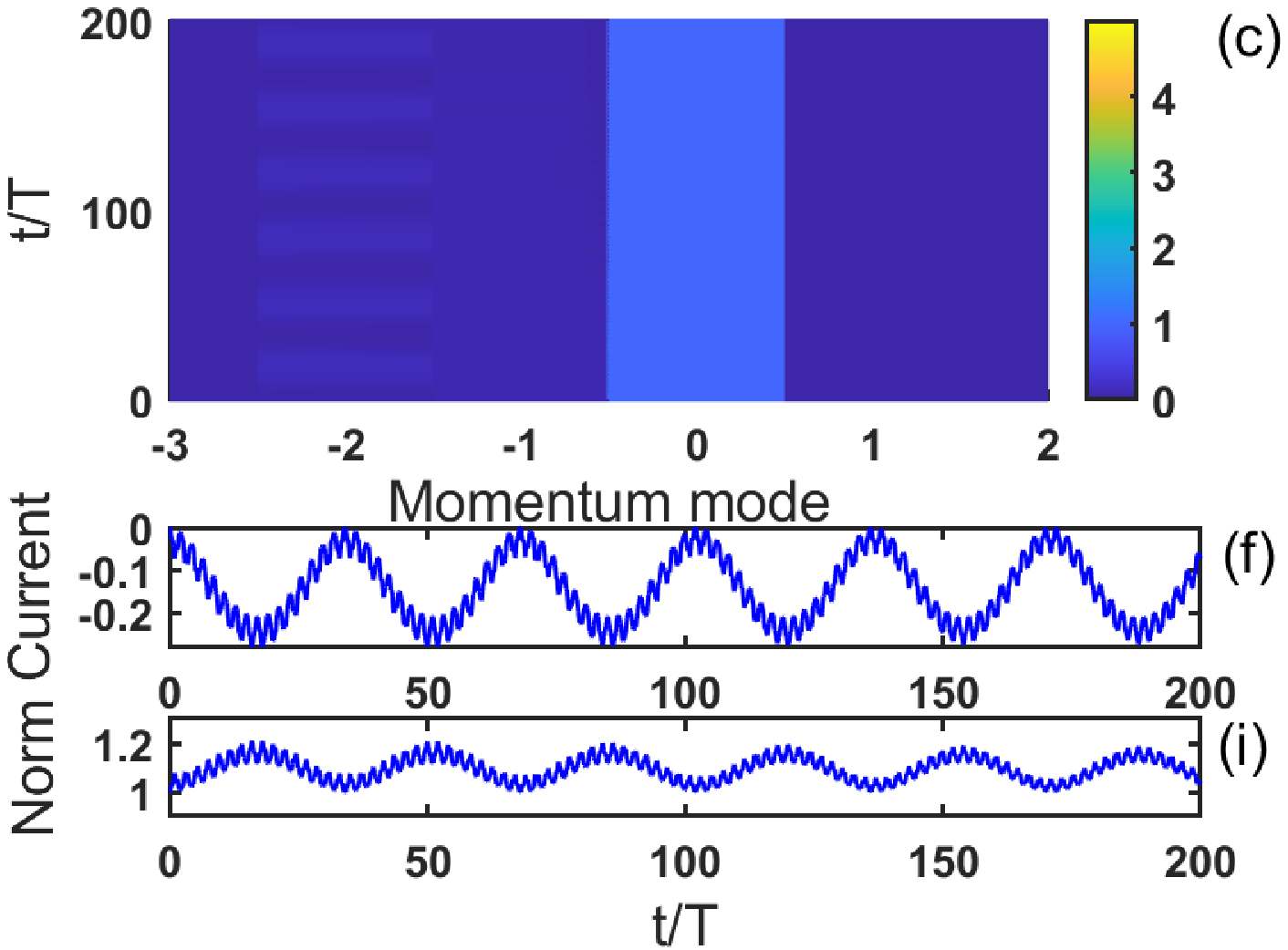}
		\label{fig10c}
	}

	\caption{The nonlinearity effects on the directed currents at EP with $K = 0.1$, $\omega = 1$, $\uplambda=1$, starting from the initial state $\ket{0}$. Top panels: time evolution of the populations in momentum space. Middle panels: time evolution of current $I(t)$. Bottom panels: time evolution of norm $\mathcal{N}(t)$.  From the left to the right column: $g=0.01, 0.1, 0.2$.}\label{fig10}
\end{figure*}
\begin{figure*}[tbp]
		\subfigure{
		\centering
		\includegraphics[scale=0.35]{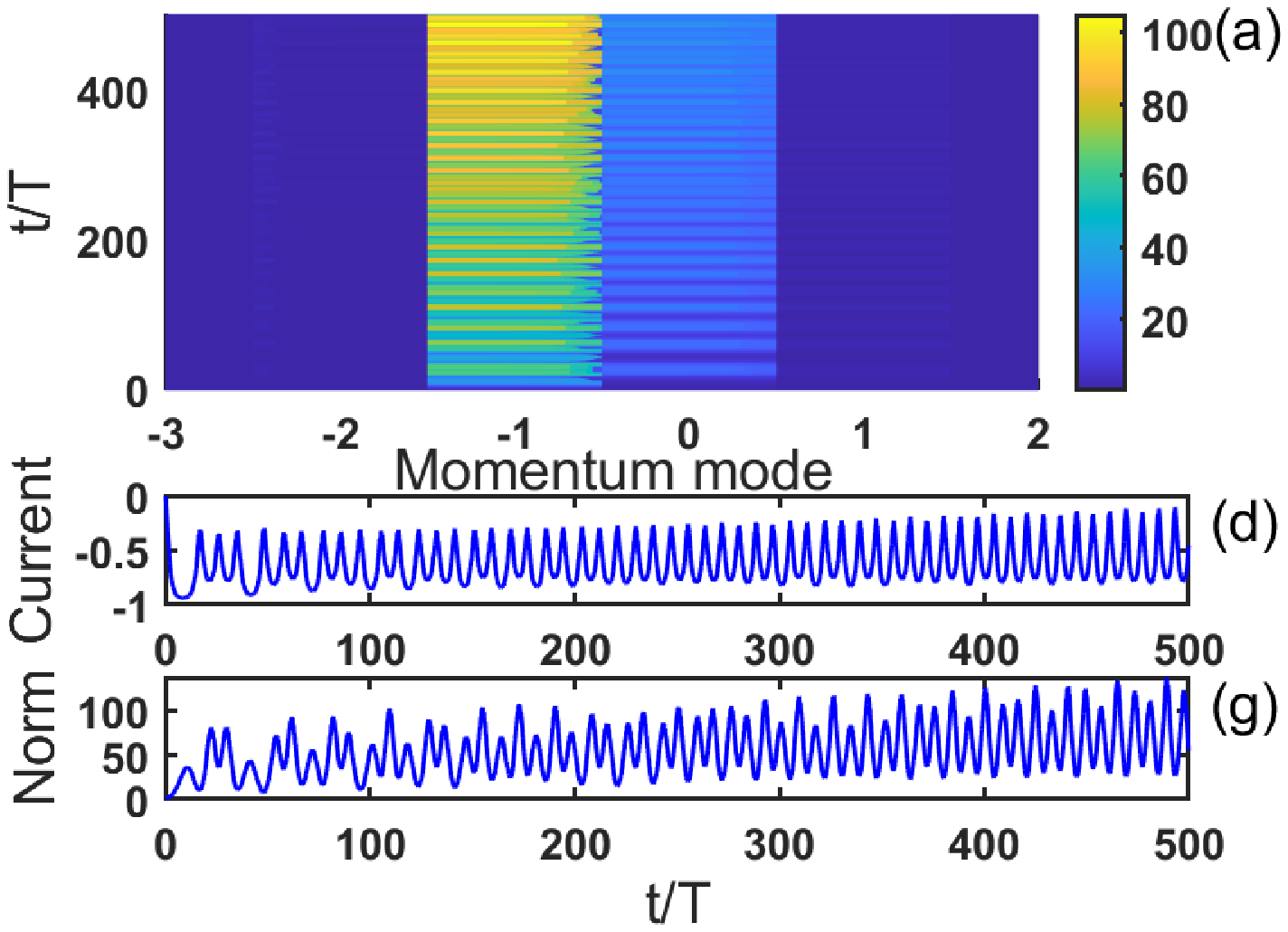}
		\label{fig11a}
	}
	\subfigure{
		\centering
		\includegraphics[scale=0.35]{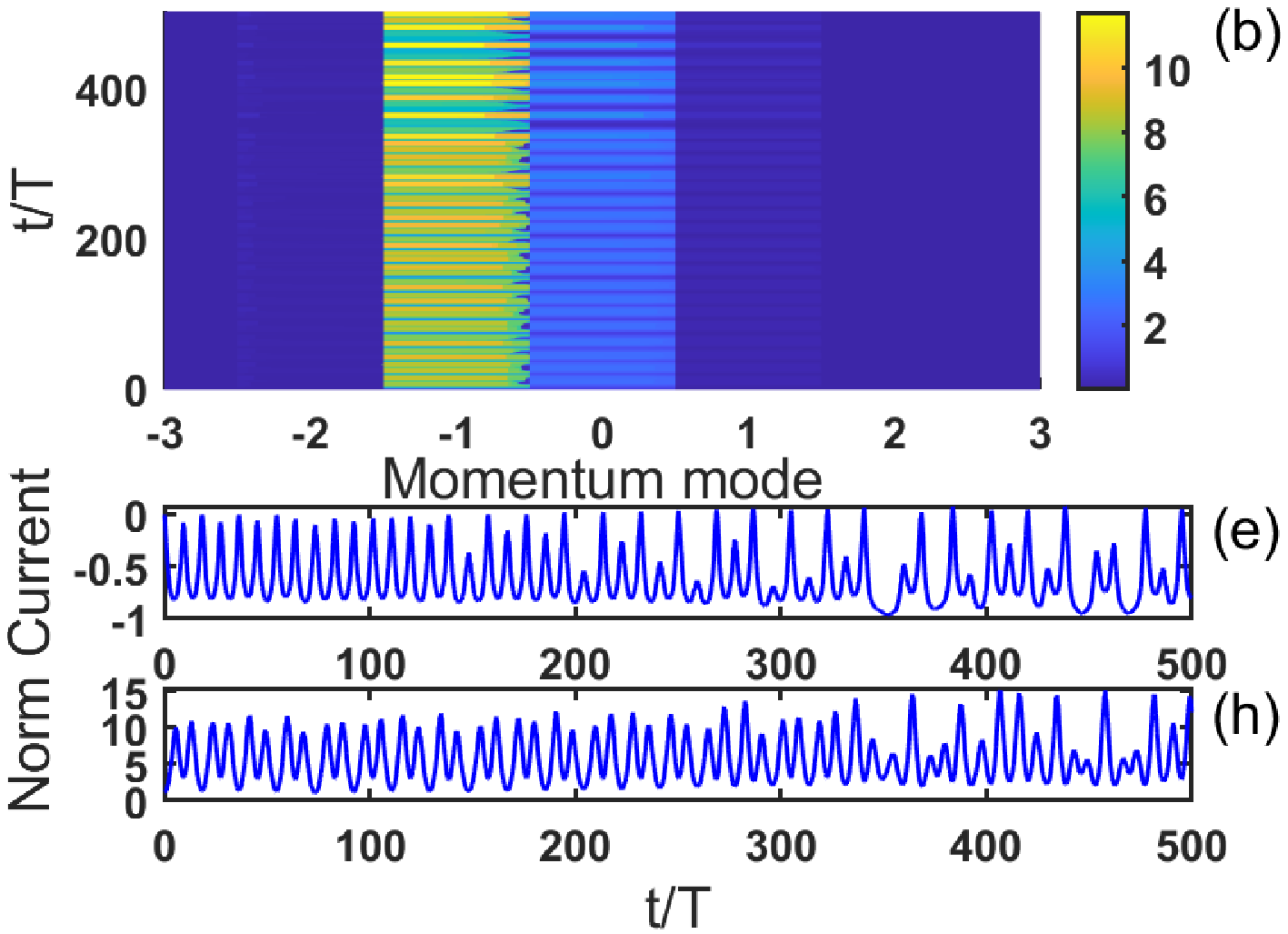}
		\label{fig11b}
	}
	\subfigure{
		\centering
		\includegraphics[scale=0.35]{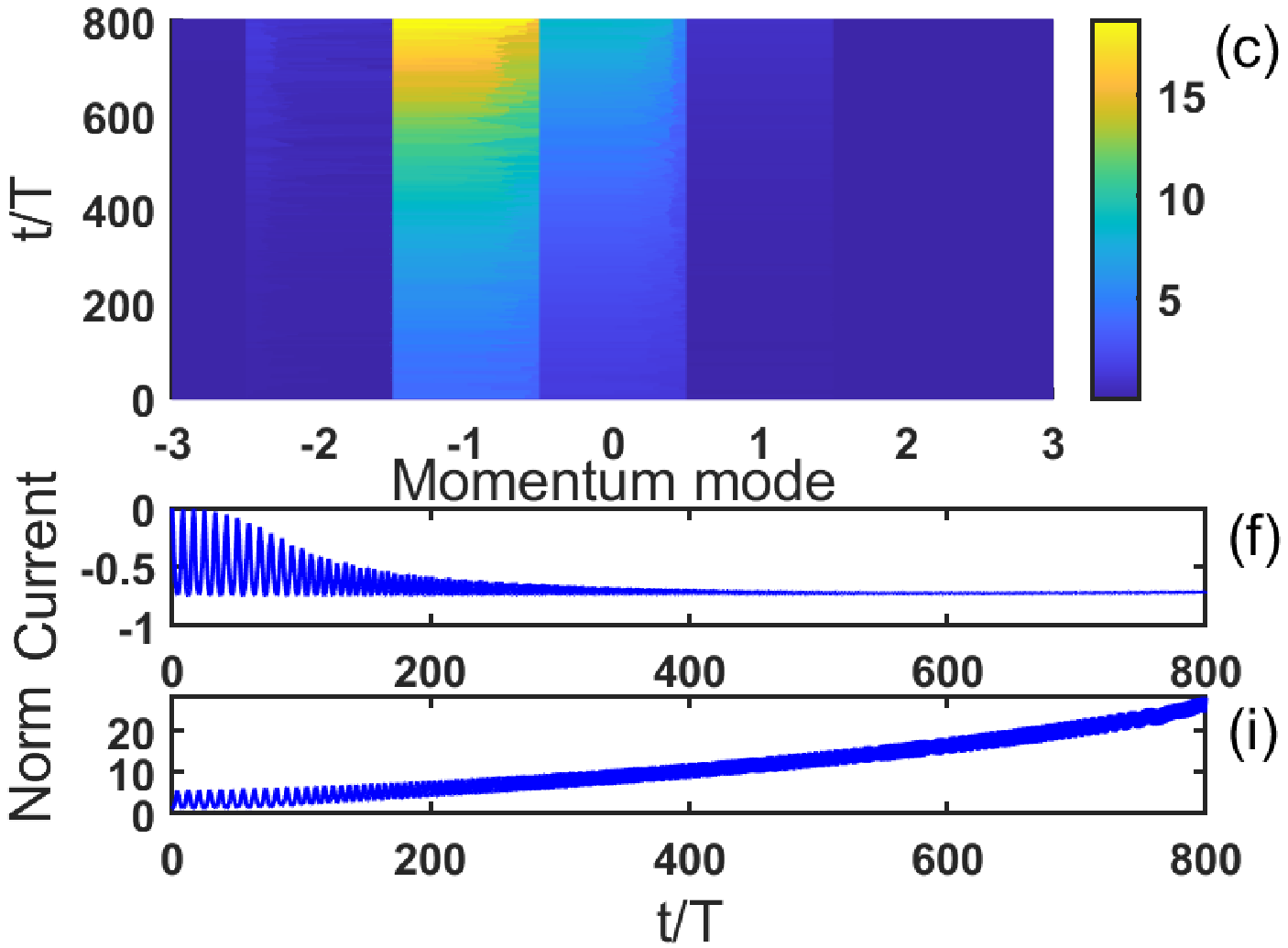}
		\label{fig11c}
	}
		\caption{The nonlinearity effects on the directed currents for the same set of parameters as in Fig. \ref{fig10}, but with $\omega=0.5$. From the left to the right column, the nonlinearity strengths are: $g=0.01, 0.1, 0.4$.}\label{fig11}
\end{figure*}
Finally, we address the impact of interatomic interactions on the current behaviours, with particular emphasis on the dynamics at the EPs. As mentioned above, for EPs in the non-interacting case, the current tends asymptotically to a constant value and the norm shows a power-law growth. First of all, we consider the parameters $\uplambda=1$ and $\omega=1$, corresponding to the EP for the non-interacting case $g=0$, which is below the symmetry breaking point and is rooted in the $\mathcal{PT}$-unbroken regime. In this case, the effects of nonlinearity on the directed current can be visualized in Fig. \ref{fig10}. In Fig. \ref{fig10}, the top row shows the time evolution of the populations in momentum space, and the middle and bottom rows show the respective currents and norms. For small $g = 0.01$ [see Fig. \ref{fig10} (a), (d) and (g)], the current oscillates periodically with time, acting like the resonant current in the $\mathcal{PT}$-unbroken regime as shown in Fig. \ref{fig3} (a). As $g$ increases (from the left column to the right column), we see that the coupling between the initial state $\ket{0}$ and other momentum states is gradually suppressed, so that the oscillatory amplitudes of the currents and norms are progressively reduced, manifesting a self-trapping phenomenon in the momentum space due to the presence of nonlinearity.

However, the dynamics are quite different for $\uplambda=1$ and $\omega = 0.5$ that correspond to the symmetry breaking point in the non-interacting case. At $g = 0.01$ [see Fig. \ref{fig11} (a), (d) and (g)], the  quadratic growth of the norm is suppressed to some extent, and the current $I(t)$ oscillates non-periodically with time. As the nonlinear strength $g$ increases to $0.1$ [see Fig. \ref{fig11} (b), (e) and (h)], the aperiodic oscillation of current and norm becomes more pronounced.  Particularly, there also exists an asymptotic current as $g$ continues to increase [see Fig \ref{fig11} (c), (f) and (i)]. The appearance of the asymptotic current is caused by the nonlinearity-induced $\mathcal{PT}$ symmetry breaking, and the population in the momentum mode $\ket{0}$ also grows with time due to the nonlinear cross-coupling of the momentum modes. This contrasts sharply with the non-interacting system, where the  population of momentum mode $\ket{0}$ remains constant because the coupling $\uplambda_-= \bra{n}\hat{V}\ket{n-1}$ vanishes at EP with $\uplambda = 1$ and the asymptotic current is induced by the EP. In the present work, we have only addressed how nonlinearity affects the current behaviour of EP in its linear counterpart. When other parameter ranges are taken into account, the corresponding current behaviour should be much richer and deserves to be explored in more detail in the future.

\section{Conclusion}\label{VII}
To conclude, we report on the non-Hermitian ratchet currents in a $\mathcal{PT}$- symmetric Floquet system with symmetric harmonic driving. In the exact $\mathcal{PT}$ phase, for a finite set of resonant frequencies, we show how a long-lasting ratchet current arises from the non-Hermiticity. We show that increasing the non-Hermitian strength can enhance the resonant currents, providing a new means of controlling the directed current for smooth continuous driving which has hitherto been studied in the Hermitian system. The resonant currents exist only when the non-Hermitian strength does not vanish, so that they are indeed a signature of non-Hermitian directed transport. The non-Hermitian resonant current reaches the largest negative value when the real and imaginary parts of the potential depth are equal, i.e. $\uplambda=1$, where the stable asymptotic current occurs owing to the exceptional points (EPs) mechanism. When $\uplambda = 1$, we find that EPs occur at the driving frequency of integer or half-integer values. More interestingly, the asymptotic current at EPs can occur below the phase transition (symmetry breaking) threshold, because such driving frequencies (e.g. $\omega=1$) of EPs, corresponding to the second-order resonance, lie in an isolated point in the $\mathcal{PT}$-unbroken regime. For weak driving, an effectively non-Hermitian three-level model is developed, which shows excellent agreement with the direct numerical results, even in the $\mathcal{PT}$-broken regime for the first-order resonance. Moreover, there also exist asymptotic currents in the broken $\mathcal{PT}$ phase, which have a different physical mechanism from the ones at the EPs. The directed currents resulting from the $\mathcal{PT}$ symmetry breaking are the consequence of the predominance of the Floquet state with the maximum imaginary value of the quasienergy over the dynamics of the non-Hermitian system. Interestingly, the momentum mode cutoff (the maximum attainable negative momentum eigenstate) of the Floquet states with the largest imaginary value of the quasienergy increases as the frequency is continuously increased, leading to an increase in the asymptotic current with the driving frequency. Finally, we consider the effect of the nonlinear interaction on the currents in the parameter configurations where the corresponding linear system lies at the EPs. It is shown that the nonlinearity can either destroy the ratchet effect due to the self-trapping in the momentum space, or induce a stable asymptotic current due to the nonlinearly induced $\mathcal{PT}$ symmetry breaking. Since the non-Hermitian Hamiltonian can be used to describe non-equilibrium relaxation problems, optical wave transport in dissipative media, exciton-polariton condensate systems, etc., our results may open new perspectives and applications of ratchet effects in a variety of non-Hermitian systems.

\acknowledgments
The work was supported by  the National
Natural Science Foundation of China (Grant No. 11975110), the Natural Science Foundation of Zhejiang Province (Grant No. LY21A050002), and Zhejiang Sci-Tech University Scientific Research
Start-up Fund (Grant No. 20062318-Y).

\section*{Appendix}\label{Appendix}
In this appendix, we give a detailed derivation of the effective three-level model (\ref{eq10})  in the main tex for the  weakly driven system under resonance. For a time-periodic Hamiltonian, there exists a complete set of solutions of the form $\ket{\phi_{\alpha}(t)}=e^{-i\varepsilon_{\alpha} t}\ket{u_\alpha(t)}$,  where the Floquet states inherit
the period of the driving, satisfying $\ket{u_\alpha(t)}=\ket{u_\alpha(t+T)}$, which can be obtained by solving the eigenvalue equation 	$[\hat{H}(t)-i\frac{\partial}{\partial t}]\ket{u_{\alpha }(t)}=\varepsilon_{\alpha }\ket{u_{\alpha }(t)}$. Note that the quasienergies and the Floquet states in the solution $\ket{\phi_{\alpha}(t)}$ are not uniquely defined. If $\ket{u_\alpha(t)}$ are the eigenstates of the eigenvalue equation with eigenvalue $\varepsilon_{\alpha}$,  then the replacement

\begin{equation}\label{eq A.1}
	\varepsilon_{\alpha m}=\varepsilon_\alpha-m\omega,\quad\ket{u_{\alpha m}(t)}=\ket{u_\alpha(t)}e^{-im\omega t},\tag{A.1}
\end{equation}
yields a new set of quasienergies and Floquet states corresponding to the same solution $\ket{\phi_\alpha(t)}=\ket{u_\alpha(t)}e^{-i\varepsilon_\alpha t}=\ket{u_{\alpha m}(t)}e^{-i\varepsilon_{\alpha m}t}$, satisfying

\begin{equation}\label{eq A.2}
	\bigg[\hat{H}(t)-i\frac{\partial}{\partial t}\bigg]\ket{u_{\alpha m}(t)}=\varepsilon_{\alpha m}\ket{u_{\alpha m}(t)}.\tag{A.2}
\end{equation}
For unperturbed system in the extended Hilbert space, Eq. (\ref{eq A.2}) reads
\begin{equation}\label{eq A.3}
	(\frac{\hat{p}^2}{2}-i\frac{\partial}{\partial t})u^0_{nm}(x,t)=\varepsilon^0_{nm}u^0_{nm}(x,t),\tag{A.3}
\end{equation}
where unperturbed Floquet states reads
\begin{equation}\label{eq A.4}
u^0_{nm}(x,t)=\braket{x,t|n,m}=\frac{1}{\sqrt{2\pi T}}\exp{(inx-im\omega t)},\tag{A.4}
\end{equation}
with the zeroth-order quasienergy
\begin{equation}\label{eq A.5}
	\varepsilon^0_{nm}=\frac{n^2}{2}-m\omega.\tag{A.5}
\end{equation}
In the following, we will concentrate on the resonance condition, defined as
\begin{equation}\label{eq A.6}
	\frac{n^2}{2}=m\omega,\tag{A.6}
\end{equation}
where the unperturbed Floquet states $\ket{n,m}$ are degenerate with $\varepsilon^0_{nm}=0$ and we expect the largest value of  the ratchet current to appear.

These degenerate resonant states are connected by the driving. The periodic driving can be expanded as Fourier components
\begin{equation}\label{eq A.7}
	\hat{V}(x,t)=\frac{K}{4}(\uplambda_-e^{ix}+\uplambda_+e^{-ix})(e^{i\omega t}-e^{-i\omega t})\tag{A.7}.
\end{equation}
We can rewrite (\ref{eq A.7}) in the extended Hilbert space using the unperturbed Floquet basis $\ket{n,m}$,
\begin{align}\label{eq A.8}
	\hat{V}=&\frac{K}{4}\sum_{n,m=-\infty}^{\infty}[(\uplambda_-(\ket{n+1,m+1}\bra{n,m}-\ket{n+1,m}\bra{n,m+1})\nonumber\\&+\uplambda_+(\ket{n,m+1}\bra{n+1,m}-\ket{n,m}\bra{n+1,m+1})].
	\tag{A.8}
\end{align}
According to the  perturbative method, the effective dynamics up to the second-order can be described by an effective $T$-matrix with the matrix elements

\begin{align}\label{eq A.9}
	\bra{n,m}\hat{T}\ket{n^\prime,m^\prime}&\simeq\bra{n,m}\hat{V}\ket{n^\prime,m^\prime}\nonumber\\&+\sum_{n^{\prime\prime},m^{\prime\prime}}\frac{\bra{n,m}\hat{V}\ket{n^{\prime\prime},m^{\prime\prime}}\bra{n^{\prime\prime},m^{\prime\prime}}\hat{V}\ket{n^\prime,m^\prime}}{-\varepsilon^0_{n^{\prime\prime},m^{\prime\prime}}}\tag{A.9}.
\end{align}

For the case $\omega = 1$, we will calculate the matrix elements restricted to the mixing of the three resonant states $\{\ket{2,2},\ket{0,0},\ket{\bar{2},2}\}$. There is no direct coupling between the initial state $\ket{0,0}$ and $\ket{2,2},\ket{\bar{2},2}$, i.e., $\bra{0,0}\hat{V}\ket{2,2}=\bra{0,0}\hat{V}\ket{\bar{2},2}=0$. The second-order mixing between $\ket{0,0}$ and $\ket{2,2},\ket{\bar{2},2}$ is generated by virtual intermediate states $\ket{1,1}$ and $\ket{\bar{1},1}$. Thus, all matrix elements $\bra{n,m}\hat{T}\ket{n^\prime,m^\prime}$ can be calculated as
\begin{equation}\label{eq A.10}
	\bra{2,2}\hat{T}\ket{0,0}=\frac{\bra{2,2}\hat{V}\ket{1,1}\bra{1,1}\hat{V}\ket{0,0}}{-\varepsilon^0_{11}}=\frac{K^2\uplambda_-}{8}\tag{A.10},
\end{equation}
\begin{equation}\label{eq A.11}
	\bra{0,0}\hat{T}\ket{2,2}=\frac{\bra{0,0}\hat{V}\ket{1,1}\bra{1,1}\hat{V}\ket{2,2}}{-\varepsilon^0_{11}}=\frac{K^2\uplambda_+}{8}\tag{A.11},
\end{equation}

\begin{equation}\label{eq A.12}
	\bra{0,0}\hat{T}\ket{\bar{2},2}=\bra{2,2}\hat{T}\ket{0,0}=\frac{K^2\uplambda_-}{8},\tag{A.12}
\end{equation}
\begin{equation}\label{eq A.13}
	\bra{\bar{2},2}\hat{T}\ket{0,0}=\bra{0,0}\hat{T}\ket{2,2}=\frac{K^2\uplambda_+}{8},\tag{A.13}
\end{equation}
with the remaining matrix elements being all zero. Restricted to these second-order matrix elements, in the space spanned only by the Floquet states $\{\ket{2,2},\ket{0,0},\ket{\bar{2},2}\}$ (using this ordering), the effective three-level model is given by the  matrix
 \begin{equation}\label{A.14}
	T\simeq\left[\begin{array}{ccc}
		0 & \frac{K^2\uplambda_-}{8} & 0 \\
		\frac{K^2\uplambda_+}{8} & 0 & \frac{K^2\uplambda_-}{8} \\
		0 & \frac{K^2\uplambda_+}{8}& 0
	\end{array}\right],\tag{A.14}
\end{equation}
which corresponds to Eq. (\ref{fig10}) in the main text.

Following the same procedure as above, under the resonance condition $\frac{n^2}{2} = m\omega$,  we can construct the effective three-level model for the $\omega = 0.5$ resonance, with the matrix defined by Eq. (\ref{eq13}) in the main text, which only involves the first-order transition between the resonant Floquet states $\{\ket{1,1},\ket{0,0},\ket{\bar{1},1}\}$.

\end{document}